\documentclass[aps,prb,twocolumn,superscriptaddress,longbibliography]{revtex4-2}

\usepackage[colorlinks=true,citecolor=blue]{hyperref}
\usepackage{graphicx}
\usepackage{amsmath}
\usepackage{amssymb}
\usepackage{appendix}
\usepackage{makecell}
\usepackage{soul}

\graphicspath{{figures/}}
\usepackage{xcolor}
\usepackage{ulem}
\usepackage{adjustbox}
\usepackage{braket}

\begin{document}

\title{Transfer learning of many-body electronic correlation entropy from local measurements}

\author{Faluke Aikebaier}
\affiliation{Department of Applied Physics, Aalto University, 00076, Espoo, Finland}
\affiliation{Computational Physics Laboratory, Physics Unit, Faculty of Engineering and Natural Sciences, Tampere University, FI-33014 Tampere, Finland}
\affiliation{Helsinki Institute of Physics P.O. Box 64, FI-00014, Finland}

\author{Teemu Ojanen}
\affiliation{Computational Physics Laboratory, Physics Unit, Faculty of Engineering and Natural Sciences, Tampere University, FI-33014 Tampere, Finland}
\affiliation{Helsinki Institute of Physics P.O. Box 64, FI-00014, Finland}

\author{Jose L. Lado}
\affiliation{Department of Applied Physics, Aalto University, 00076, Espoo, Finland}

\begin{abstract}
Characterizing quantum correlations in many-body systems is essential for understanding the emergent phenomena in quantum materials. The correlation entropy serves as a key metric for assessing the complexity of a quantum many-body state in interacting electronic systems. However, its determination requires the measurement of all single-particle correlators across a macroscopic sample, which can be impractical. Machine learning methods
have been shown to allow learning the correlation entropy from a reduced set of measurements, 
yet these methods assume
that the targeted system is contained in the set of training
Hamiltonians.
Here we show that a transfer learning strategy 
enables correlation entropy learning
from a reduced set of measurements
in variants of Hamiltonians never considered in the training set.
We demonstrate this transfer learning methodology in a wide variety of interacting models including
local and non-local attractive and repulsive many-body interactions, 
long range hopping, doping, magnetic field
and spin-orbit coupling.
Furthermore, we show that this transfer learning methodology allows detecting phases in quantum many-body systems beyond those present in the training set.
Our results demonstrate that 
correlation entropy learning
can be potentially 
performed experimentally without requiring training
in the experimentally realized Hamiltonian.
\end{abstract} 

\maketitle

\section{Introduction}

Quantum correlations between interacting particles are essential for understanding complex behaviors in many-body quantum systems~\cite{Fulde1995,Paschen2020}, and reflect the intrinsic complexity of these states~\cite{Anderson1987,Savary2016,Keimer2017,Lau2020}. Traditional entanglement measures, such as the von Neumann and R\'enyi entropies, are commonly used to quantify quantum correlations~\cite{Eisert2010,Amico2008,Horodecki2009}. However, these measures can yield non-zero values even for non-interacting systems~\cite{Gu2004,Korepin2004,Deng2006,Iemini2015}. A more direct approach is to use the von Neumann entropy of the one-particle density matrix (ODM), which vanishes in the noninteracting limit and more accurately reflects the correlations arising from many-body interactions~\cite{Wang2007,Tichy2011,Hofer2013,Esquivel2015}. Nevertheless, directly measuring such correlation entropy remains challenging, as it requires detailed particle-particle correlators of systems, which is especially impractical for large systems ~\cite{Esquivel1996,Gersdorf1997,Ziesche1997,Huang2006,BenavidesRiveros2017,Ferreira2022}.

\begin{figure}[t!]
\includegraphics[width=1.0\linewidth]{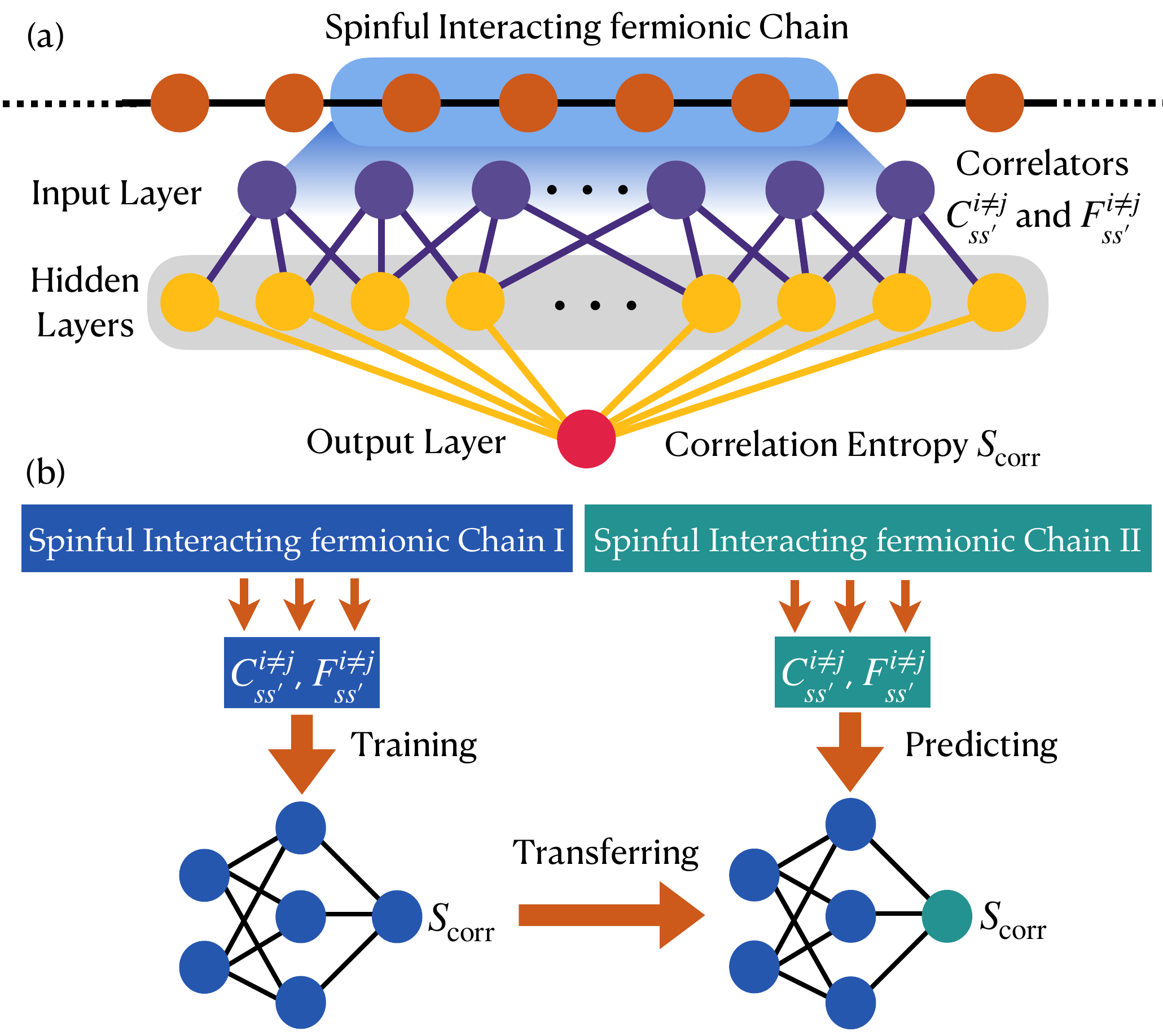}%
\caption{\label{fig:schematics} Schematics of the model. (a) The neural network algorithms to extract quantum correlations from local measurable correlators. (b) A neural network algorithm trained on correlators extracted from one interacting fermionic chain is used to predict the correlation entropy of another interacting fermionic chain with a different parameter regime.  }
\end{figure}

Machine learning strategies enable the extraction of non trivial properties of quantum many-body states requiring solely specific observables\cite{PhysRevB.97.115453,PhysRevB.110.075402,Yu2022,Kming2021,PhysRevLett.125.127401,PhysRevB.97.205110,RodriguezNieva2019,PhysRevApplied.20.044081,PhysRevApplied.20.044081,2024arXiv240506688L,2024arXiv241207666L,2024arXiv241212019S,PhysRevA.105.023302,PhysRevB.111.014501,PhysRevApplied.20.024054,Khosravian2024,Schleder2019,Bedolla2020,PhysRevB.102.054107,10.21468/SciPostPhysCore.6.2.030,PhysRevB.109.195125}.
In the case of the fermionic correlation entropy, machine learning
allows inference from a handful of local measurements without requiring access to the full set of correlators~\cite{10.21468/SciPostPhysCore.6.2.030,PhysRevB.109.195125}. This approach significantly simplifies the process, offering a more efficient way to determine the correlation entropy while retaining accuracy, thereby serving as a valuable tool for bridging the gap between theory and experiment.
This strategy is analogous to machine learning-enabled quantum tomography methods~\cite{Torlai2023,PRXQuantum.4.040337,PRXQuantum.3.020344,SurawyStepney2022projectedleast,PRXQuantum.5.040306},
which allow characterizing exponentially large quantum many-body states from a properly chosen set of measurements.
However, a potential limitation for machine learning correlation entropy
stems from the fact that the training data must be generated with specific quantum models,
which may or may not correspond to the system observed experimentally.
Transfer learning~\cite{https://doi.org/10.48550/arxiv.1808.01974,Iman2023}
tackles the question of whether machine learning algorithms can be
extended beyond their simplified and highly controlled training set,
a common scenario for machine learning methodologies in physics,
ranging from soft matter~\cite{Goetz2022,Wang2019}, electronic structure~\cite{Rowe2020,Keith2021} 
and many-body physics~\cite{PhysRevE.101.053301,Sayyad2024}.
This motivates the question of whether an 
algorithm trained with models that do not correspond
to the experimental one would enable learning the correlation entropy.

Here we demonstrate
a machine learning 
strategy to extract the correlation entropy
of a many-body system, where the training is performed 
on a different class of Hamiltonians
from the tested ones.
Our methodology enables 
predicting the correlation entropy of many-body models
never seen during the training.
This enhancement broadens the applicability of machine learning
methodologies to learn the correlation entropy, 
allowing for the characterization of quantum correlations in a wider variety of electronic systems without the need for potential retraining. To achieve this goal, we use a transfer strategy as shown in Fig.~\ref{fig:schematics}(b), to create a robust algorithm that maintains high predictive accuracy even when applied to different unseen systems. Furthermore, we also show the robustness of such models to potential noise in the expectation values used, 
ensuring that the model remains reliable in practical conditions. 
The paper is organized as follows. In Sec.~\ref{sec:corr_entropy}, we introduce the concept of correlation entropy and outline 
the variants of Hamiltonians considered. In Sec.~\ref{sec:transfer_learning}, we provide a detailed discussion of the transfer learning framework and evaluate the predictive performance of the neural network algorithms. Sec.~\ref{sec:effect_noise} examines the impact of noise on the accuracy of the model. Finally, we present our conclusions in Sec.~\ref{sec:conclusion}. 

\section{Correlation Entropy\label{sec:corr_entropy}}
In the presence of many-body interactions, quantum many-body states may strongly depart from conventional mean-field
Hartree-Fock states. As a measure of such deviation, the correlation entropy is defined in terms of the correlation matrix\cite{PhysRev.97.1474,Wang2007,Tichy2011,PhysRevB.89.085108,PhysRevB.92.075132,PhysRevResearch.6.023178,PhysRevB.103.235166,PhysRevB.108.205120,PhysRevResearch.6.L012052,PhysRevResearch.6.013060,10.21468/SciPostPhysCore.6.2.030,2024arXiv240917202L,2024arXiv241209576G}. 
The correlation matrix constitutes of particle-particle correlators 
\begin{equation}\label{eq:ppcorrelator}
C_{ij}^{ss'}=\left\langle\Psi_0\right.\vert c_{is}^{\dagger}c_{js'}\left.\vert\Psi_0\right\rangle, 
\end{equation}
where $|\Psi_0\rangle$ is the many-body ground state, $c_{is}$, $c^\dagger_{is}$ are the annihilation and creation operators at site $i$ and spin $s$. Due to Fermi statistics, the eigenvalues $\alpha_\nu$ of the correlation matrix satisfy $0\leq\alpha_\nu\leq 1$. This allows to define the correlation entropy as 
\begin{equation}\label{eq:CorrelationEntropy}
    S_{\rm{corr}}=-\sum_{\nu=1}^{2N}\alpha_\nu\log \alpha_\nu,
\end{equation}
where $N$ is the size of the system. The correlation entropy is nonnegative, scales linearly with the system size for
ground states of local Hamiltonians~\cite{Eisert2010},
and is exactly zero for the ground state
of non-interacting fermionic systems. 
The correlation entropy is a measure of how far away a given state,
and in particular the ground state of a many-body Hamiltonian,
is from the set of fermionic Hartree-Fock product states. 

We consider various one-dimensional extended models of spinful interacting fermionic chains of the form
\begin{equation}\label{eq:Hamiltonians}
    H=H_0+H_p,
\end{equation}
where 
\begin{equation}\label{eq:pureHubbardModel}
\begin{split}
H_0=&-t\sum_{j,s}\left(c_{j,s}^{\dagger}c_{j+1,s}+h.c. \right)\\
&+U\sum_{j}\left(n_{j,\uparrow}-\frac{1}{2} \right)\left(n_{j,\downarrow}-\frac{1}{2} \right)
\end{split}
\end{equation}
is the half filled pure Hubbard model and
$p=\mu,t^{\prime},V,h,\lambda_R,J$ represents chemical potential, next nearest-neighbour hopping, nearest-neighbour interaction, magnetic field in $z$ direction, Rashba spin-orbit interaction, and spin-spin interaction, respectively. The number operator is defined as $n_{js}=c^{\dagger}_{js}c_{js}$ and $s=\uparrow/\downarrow$ is spin. Then the last term in the Hamiltonian can be specifically written as
\begin{equation}
    H_{\mu}=-\mu\sum_{j}\left(n_{j,\uparrow}+n_{j,\downarrow}\right)
\end{equation}
\begin{equation}
    H_{t^{\prime}}=t^{\prime}\sum_{j,s}\left(c_{j,s}^{\dagger}c_{j+2,s}+h.c. \right)
\end{equation}
\begin{equation}
    H_V=V\sum_{j}\left(n_j-1 \right)\left(n_{j+1}-1 \right)
\end{equation}
\begin{equation}
    H_{h}=-\frac{h}{2}\sum_{j}\sigma^z_{s,s'} c^\dagger_{j,s} c_{j,s'}
\end{equation}
\begin{equation}
    H_{\lambda_R}=i\lambda_R\sum_{\langle jk \rangle ss'}c_{j,s}^{\dagger}\left(\boldsymbol{\sigma}_{ss'}\times\boldsymbol{d}_{jk} \right)_zc_{k,s'}+{\rm{h.c.}}
\end{equation}
\begin{equation}
    H_{J}=J\sum_{i}\boldsymbol{S}_i\cdot\boldsymbol{S}_{i+1},
\end{equation}
where $\boldsymbol{d}_{jk}= \boldsymbol{r}_{j} - \boldsymbol{r}_{k}$ with $\boldsymbol{r}_{j}$ the location of site $j$, $\langle jk \rangle$ denotes first neighboring sites, $\boldsymbol{\sigma}=(\sigma_x,\sigma_y,\sigma_z)$ are the spin Pauli matrices, and $\boldsymbol{S}_i$ is the spin operator as site $i$. Distinct terms in the single-particle Hamiltonian are responsible for breaking different symmetries. In the half-filled Hubbard model, for instance, next-nearest-neighbor hopping breaks chiral symmetry, the chemical potential breaks electron–hole symmetry, and Rashba spin–orbit coupling breaks spin rotational symmetry. In realistic materials, it is often the case that only a reduced set of these perturbations is dominant, thereby placing the system effectively within one of the specific symmetry classes considered here.

\begin{figure*}
\includegraphics[width=1.0\textwidth]{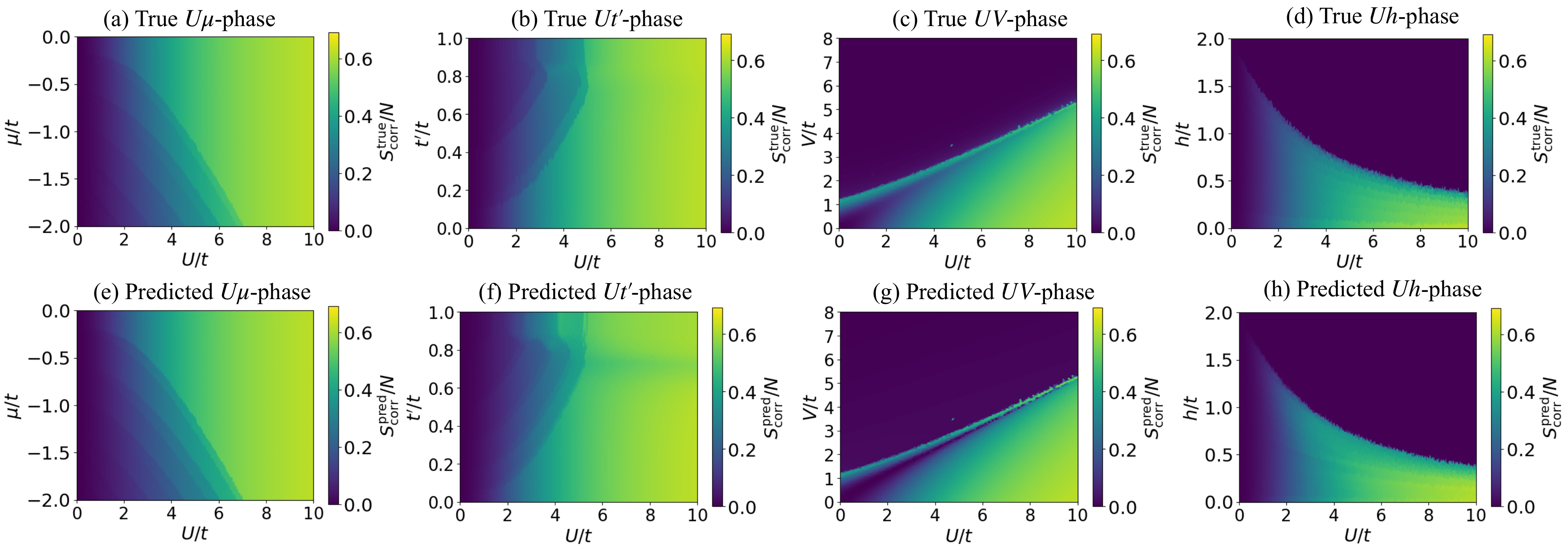}%
\caption{\label{fig:phase_diagrams} Phase diagrams of the correlation entropy for several interacting models. The upper panels show the true phase diagrams of (a) $\mu U$, (b) $t^{\prime}U$, (c) $VU$, (d) $hU$. The lower panels show the predicted phase diagrams of the model trained with the parameter regime $H=H(t,U,\mu>0)$ in the corresponding order. The fidelity of these predictions are all $\mathcal{F}\simeq1$. The phases identified in the diagrams
can be compared to those from Refs.~\cite{MOTT1968,Nishimoto2008,Yu2016,vanDongen1997}.}%
\end{figure*}

The phases of selected extended models based on the correlation entropy are shown in Fig.~\ref{fig:phase_diagrams}(a-d). The Hamiltonians are solved with open boundary conditions by using the tensor-network formalism~\cite{PhysRevLett.69.2863,2020arXiv200714822F,ITensor,DMRGpyLibrary} on a one-dimensional fermionic chain with 32 spinful sites, and the correlation entropy is calculated by Eq.~\eqref{eq:CorrelationEntropy}. We use 32 spinful sites as the correlation entropy in this case is already independent from the system size~\cite{10.21468/SciPostPhysCore.6.2.030}.  As shown, for vanishing interactions $U=0$, the correlation entropy vanishes. The correlation entropy also vanishes for quantum states that do not support sizable many-body quantum entanglement even at finite many-body interactions~\cite{Gu2004,Deng2006,Huang2006,Glocke2007}. For quantum states with sizeable entanglement, the correlation entropy saturates to the value $\ln 2$ per site~\cite{Ziesche1997}. 
For a system
with $N$ spinful sites, the correlation matrix has dimensions $2N\times 2N$. For a large system, measuring the correlation entropy faces significant challenges as it would require knowledge of the full correlation matrix~\cite{Bergschneider2019}. 

While the formal definition of the correlation entropy requires
the full particle-particle correlation matrix (Eq.~\eqref{eq:ppcorrelator}), 
machine learning methodologies
offer a workaround only employing a 
small number of local measurements\cite{10.21468/SciPostPhysCore.6.2.030,PhysRevB.109.195125}.
The information lost by only considering local measurements
is compensated by including both 
local particle-particle and density-density correlators\cite{10.21468/SciPostPhysCore.6.2.030,PhysRevB.109.195125}
\begin{equation}\label{eq:nncorrelator}
    f_{ij}^{ss'}=\left\langle\Psi_0\right.\vert n_{is}n_{js'} \left.\vert\Psi_0\right\rangle
\end{equation}
extracted from the same sites~\cite{10.21468/SciPostPhysCore.6.2.030}. These two types of correlators offer complementary insight into the many-body state~\cite{DelMaestro2022}. {While particle-particle correlators directly encode correlation information relevant to the definition of correlation entropy, density-density correlators capture complementary features, including beyond Hartree-Fock corrections
reflecting complementary information to non-local correlators.}  
The core concept of developing a supervised algorithm is to reverse-engineer the correlation entropy from a limited set of these local correlators~\cite{10.21468/SciPostPhysCore.6.2.030,PhysRevB.109.195125}. A supervised algorithm on a one-dimensional generalized model of an interacting spinful fermionic chain allows to accurately predict the correlation entropy across different parameter subspaces of the chain depicted in Fig.~\ref{fig:schematics}(a). However, for correlators outside the parameter regime of the training data, 
the algorithm must be trained again in general.

\section{Transfer Learning\label{sec:transfer_learning}}
In our work, we focus on whether an algorithm trained on one variant of the extended Hubbard model can accurately predict the correlation entropy in a different variant, thus realizing a transferable
algorithm between different quantum many-body systems~\cite{PhysRevE.101.053301,Sayyad2024}.
To demonstrate our transfer learning strategy, we consider the schematic shown in Fig.~\ref{fig:schematics}(b). By training a neural-network algorithm on correlators from one fermionic chain, we aim to predict the correlation entropy of another fermionic chain in a different parameter regime. 
However, carrying out this procedure directly with
the correlation matrices in Eq.~\eqref{eq:ppcorrelator} and
Eq.~\eqref{eq:nncorrelator} leads to a failure of the
transfer learning.
We can overcome this limitation by enforcing using only
correlators between different sites, and redefining the density-density correlators as
\begin{equation}\label{eq:nncorrelator2}
    F_{i\neq j}^{ss'}=\left\langle\Psi_0\right.\vert n_{is}n_{js'}\vert\left.\Psi_0\right\rangle-\left\langle\Psi_0\right.\vert n_{is}\vert\left.\Psi_0\right\rangle\left\langle\Psi_0\right.\vert n_{js'}\vert\left.\Psi_0\right\rangle.
\end{equation}

It is worth noting that the inclusion of density-density correlators enables the algorithm to observe
corrections beyond mean-field that can be rationalized as follows. 
Formally, the correlation entropy can be computed solely from knowing all particle-particle correlators, without density-density correlators. Nevertheless, since our approach leverages only local correlators, the inclusion of local density-density correlators provides instrumental complementary information to extract the correlation entropy. This can be easily rationalized from Wick's theorem for non-interacting electrons.
For a ground state described by a Hartree Fock state $\Psi_{HF}$, Wick's theorem allows expressing the density-density correlators
as a function of the particle particle correlators as
$
\langle c^\dagger_{i,s} c_{i,s} c^\dagger_{j,s'} c_{j,s'} \rangle_{\Psi_{HF}} =
\langle c^\dagger_{i,s} c_{i,s} \rangle_{\Psi_{HF}} \langle c^\dagger_{j,s'} c_{j,s'} \rangle_{\Psi_{HF}}
-
\langle c^\dagger_{i,s} c_{j,s'} \rangle_{\Psi_{HF}} \langle c^\dagger_{j,s'} c_{i,s} \rangle_{\Psi_{HF}}
$
As a result, the following quantity enables to directly distinguish between a Hartree-Fock state, and a 
many-body state with electronic entanglement for a generic wavefunction $\Psi$
$
\Xi =
\langle c^\dagger_{i,s} c_{i,s} c^\dagger_{j,s'} c_{j,s'} \rangle_{\Psi}
-
\langle c^\dagger_{i,s} c_{i,s} \rangle_{\Psi} \langle c^\dagger_{j,s'} c_{j,s'} \rangle_{\Psi}
+
\langle c^\dagger_{i,s} c_{j,s'} \rangle_{\Psi} \langle c^\dagger_{j,s'} c_{i,s} \rangle_{\Psi}
$.
It is worth noting that a non-zero $\Xi$ in the local correlators directly implies that the systems features finite
electronic correlation. As a result, information about local density-density and particle-particle correlators
alone enables identifying the presence of electronic correlation entropy, yet without providing a quantitative
estimate. Our machine learning algorithm leverages this information to provide a quantitative extraction of
the correlation entropy.

\begin{figure}[t!]
\includegraphics[width=1.0\linewidth]{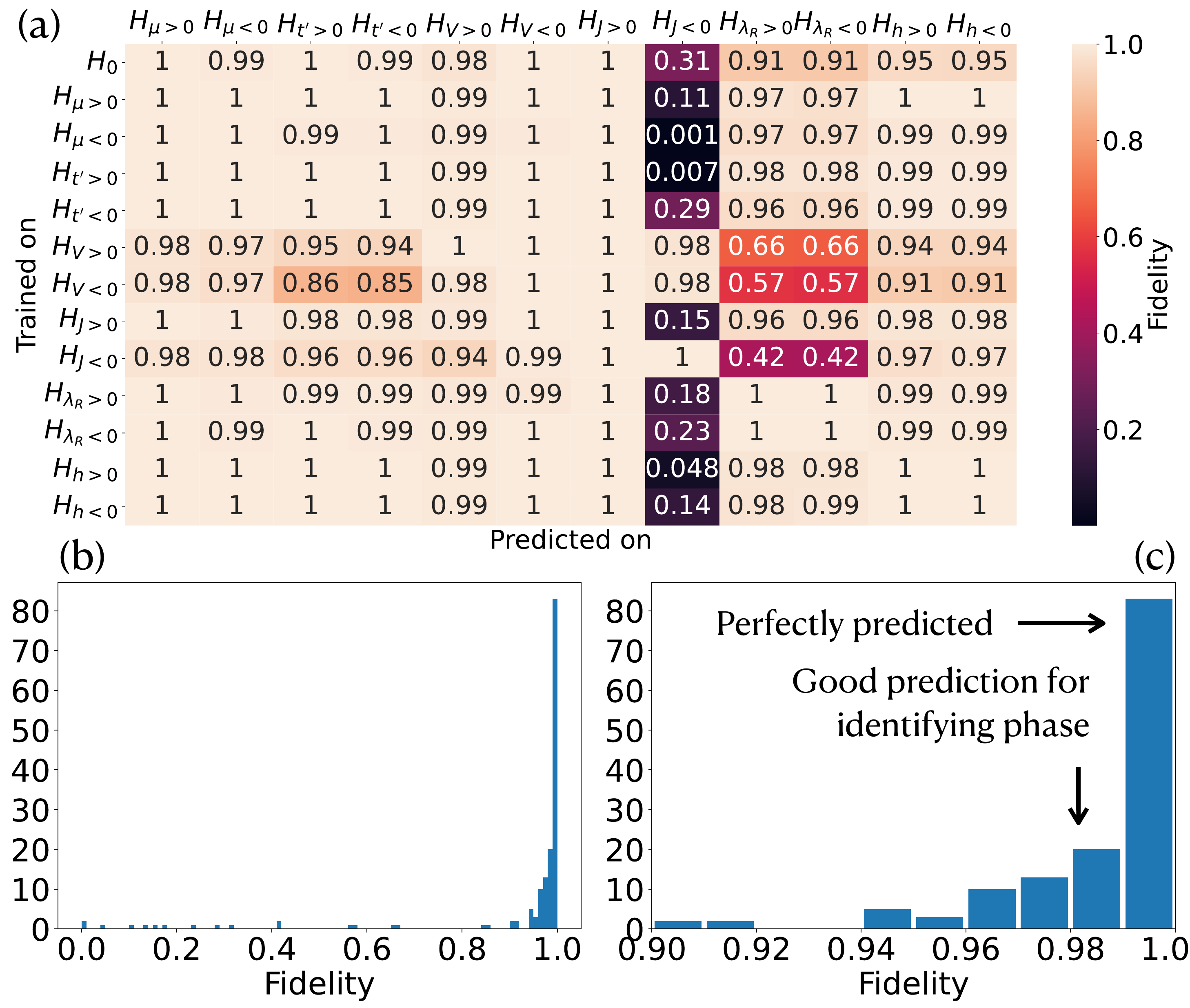}%
\caption{\label{fig:fidelity} (a) Fidelity table for the transfer learning. On each row, a neural-network algorithm trained on the Hamiltonian parameters labeled on the left, and predict on the dataset with the Hamiltonian parameters labeled on the top. (b) Histogram of (a) based on the fidelity value, and (c) shows a smaller fidelity value range. }
\end{figure}

Thus the correlators are extracted from the modified correlation matrices $\Lambda_{ij}^{ss'}=\{C_{i\neq j}^{ss'},F_{i\neq j}^{ss'}\}$. In order to reduce the influence of finite-size effect, the correlators are extracted from the central four sites of the fermionic chain. 

In order to examine the accuracy of the model on unseen Hamiltonian parameter regime, we use the same neural-network structure for all the possible algorithms, with three hidden layers and 1024 nodes on each layer. 
We characterize the quality of the prediction by means of the prediction
fidelity~\cite{Khosravian2024,PhysRevB.109.195125}, defined as
\begin{equation}\label{eq:Fidelity}
\mathcal{F} = 
    \frac{\big\vert \left\langle S_{\rm{corr}}^{\rm{pred}}\cdot S_{\rm{corr}}^{\rm{true}}\right\rangle -\left\langle S_{\rm{corr}}^{\rm{pred}}\right\rangle \cdot \left\langle S_{\rm{corr}}^{\rm{true}}\right\rangle \big\vert}{\sqrt{ \left[\big\langle \left(S_{\rm{corr}}^{\rm{true}}\right)^2\big\rangle-\left\langle S_{\rm{corr}}^{\rm{true}}\right\rangle^2\right] \left[\big\langle \big(S_{\rm{corr}}^{\rm{pred}}\big)^2\big\rangle-\big\langle S_{\rm{corr}}^{\rm{pred}}\big\rangle^2\right] }}
\end{equation}
and standard deviation
\begin{equation}\label{eq:standarddeviation}
    \sigma=\sqrt{\frac{\sum_{i}^{n}\left(S_{\rm{i,corr}}^{\rm{pred}}-\left\langle S_{\rm{corr}}^{\rm{pred}} \right\rangle \right)^2}{n}},
\end{equation}
where $i$ runs over all the trained neural-network algorithms, as the metrics for the algorithms. {Here $S_{\rm{i,corr}}^{\rm{pred}}$ refers to the output of the $i$-th model for a given data point, and the average correlation entropy $\left\langle S_{\rm{corr}}^{\rm{pred}} \right\rangle$ is taken over these multiple predictions. The standard deviation $\sigma$ is computed across these independent model outputs to quantify the model-to-model consistency and estimate uncertainty in the prediction.}

The transferability of the neural-network algorithms can be examined by training a neural-network algorithm within one Hamiltonian parameter regime, namely, one extended version of Hamiltonian in Eq.~\eqref{eq:Hamiltonians}, and predicting on another. For each extended version of Hamiltonian, the neural-network algorithm is trained on examples randomly generated within the parameter regime. Note that for each extended version of the Hamiltonian, the quantum many-body phases are intrinsically linked to the range of tight-binding parameters and the corresponding correlation entropy. In other words, phases that lie outside these parameter ranges cannot be accessed. The accuracy of the prediction is represented by the fidelity in Eq.~\eqref{eq:Fidelity}, and the results are shown in Fig.~\ref{fig:fidelity}(a) as a fidelity table. For simplicity, we only consider repulsive interaction $U>0$ and denote each extended versions of Hamiltonian by the second term $H_p$ in Eq.~\eqref{eq:Hamiltonians}, including both positive and negative values of the tight-binding parameters. 

The fidelity table shows that neural-network algorithms can accurately predict correlation entropy as evidenced by the high values in the table. Trivially, high accuracy is maintained within their Hamiltonian parameter subspaces and when changing the sign of Hamiltonian parameters, as the correlators used in training for these cases are quite similar.
More importantly, the algorithms generally perform well in predicting the correlation entropy across different Hamiltonian parameter regimes. However, notable exceptions include systems with ferromagnetic phase, specifically $H_{J<0}$, and systems with Rashba spin-orbit interaction $H_{\lambda_R}$.  
These phases exhibit distinct patterns in their correlators that differ significantly from those in the majority of extended Hubbard models described in Eq.~\eqref{eq:Hamiltonians}. To illustrate clearly, we take an algorithm trained with $H_{\mu>0}$, and show the best prediction on four cases in Eq.~\eqref{eq:Hamiltonians}. The corresponding phase diagrams are shown in Fig.~\ref{fig:phase_diagrams}(e-h). One can see that, the transfer learning approach can identify phases in quantum many-body systems that were not included in the training set. 

\begin{figure}[t!]
\includegraphics[width=1.0\linewidth]{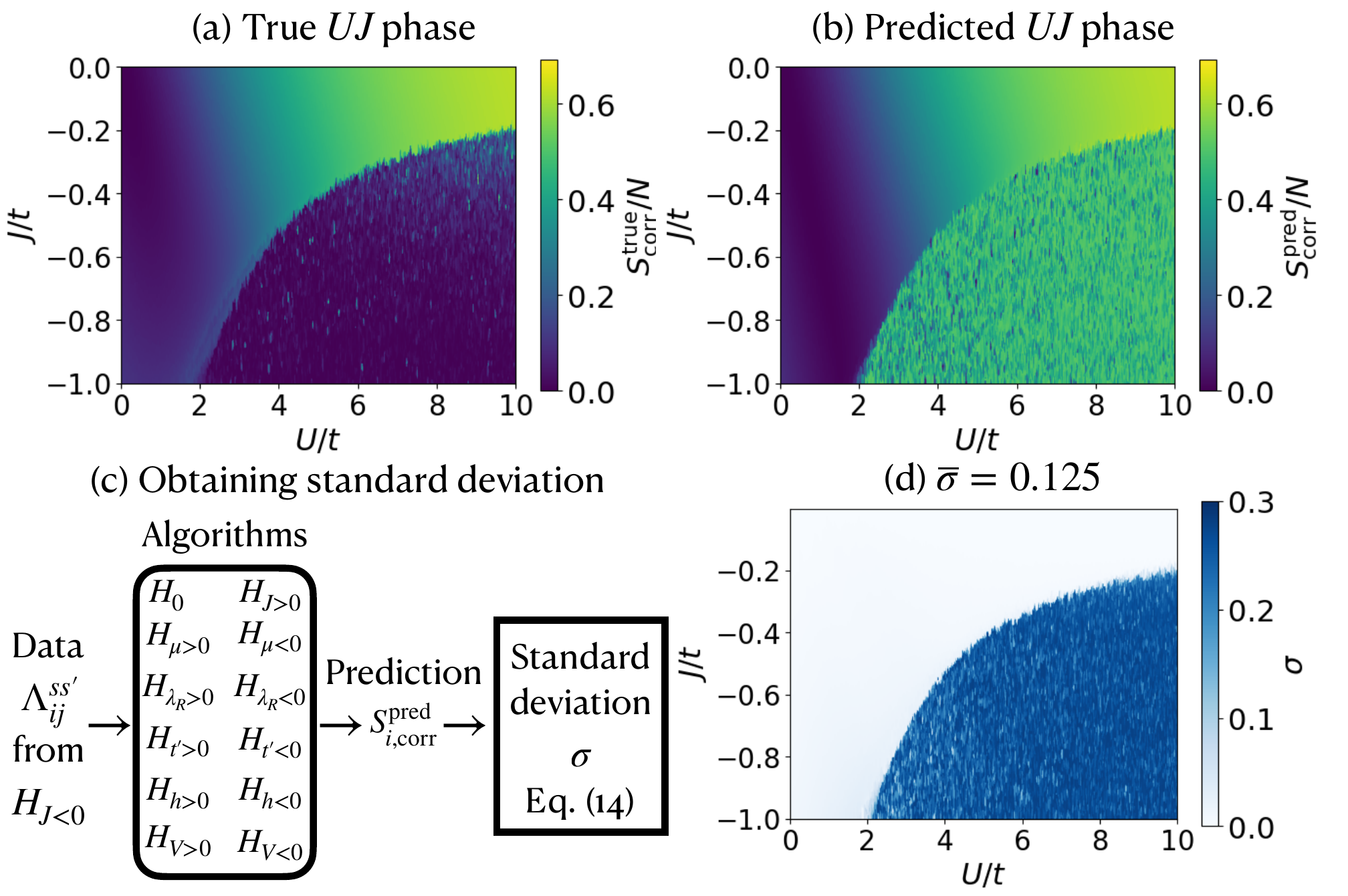}%
\caption{\label{fig:standard_deviation} (a, b) The true and the predicted phase diagrams of $H_{J<0}$, with the algorithm trained on the data extracted from $H_{\mu>0}$. (c, d) Schematic illustration and the result of the standard deviation of the datasets  $H_{J<0}$ predicted by all the trained algorithms, respectively. The total standard deviation $\overline{\sigma}$ is given as the title of (d). }%
\end{figure}

The difficulty of algorithms in predicting certain data sets can be further quantified by examining the standard deviation in the prediction. It reflects the inconsistency in the performance of the algorithm in these different parameter regimes. Specifically, we can apply all trained algorithms separately to each data set with unseen parameter regimes and analyze the resulting phase diagrams in terms of the standard deviation and the total standard deviation. The results of the ferromagnetic state is shown in Fig.~\ref{fig:standard_deviation}. 
As shown in Fig.~\ref{fig:standard_deviation}, the predictions exhibit wide deviations in some parameter ranges representing rather distinct quantum phases compared to the trained data sets. In the case of the ferromagnetic phase induced by the spin-spin interaction, the algorithms struggle to accurately predict the true values of the correlation entropy. However, they can still distinguish the ferromagnetic phase from the antiferromagnetic phase.  

A statistical analysis of the transferability of the neural-network algorithms in terms
of the fidelity table is shown in Fig.~\ref{fig:fidelity}(b,c). It is observed that half of the datasets are accurately predicted with $\mathcal{F}=1$. Combining with the results represented by standard deviation in Fig.~\ref{fig:standard_deviation}, the rest of the predictions are sufficiently accurate to correctly identify the phases.
We observe that the predicted correlation entropy is reliable as long as the quantum order of the target system is not drastically different from that of the system on which the algorithm was trained,
even if its parent Hamiltonian is completely different. 
Since most of the algorithms in our study were trained on systems
featuring antiferromagnetic quantum correlations, 
they tend to perform poorly when predicting systems
with ferromagnetic fluctuations and 
systems with Rashba spin-orbit interaction.

\section{Noise and robustness\label{sec:effect_noise}}
In actual experimental scenarios, measured correlators feature some unavoidable
noise. Therefore, it is instrumental to examine the robustness of the algorithms in the presence
of noisy expectation values, ensuring that the trained algorithm 
would be reliable under experimental conditions.
The robustness of the neural network model can be examined by introducing random noise in the correlators. 
For each example in the training data, the random noise is included as follows,
\begin{equation}\label{eq:random_noise}
    \Lambda_{ij}^{ss'}=\Lambda_{ij}^{ss',0}+\chi_{ij}^{ss'},
\end{equation}
where $\Lambda_{ij}^{ss',0}=\{C_{i\neq j}^{ss'},F_{i\neq j}^{ss'}\}$ is the original correlators, and $\chi_{ij}^{ss'}$ the random noise between $[-\omega,\omega]$, and $\omega$ is the noise amplitude. To quantify the impact of noise, we evaluate the fidelity for different levels of noise.

\begin{figure}[t!]
\includegraphics[width=1.0\linewidth]{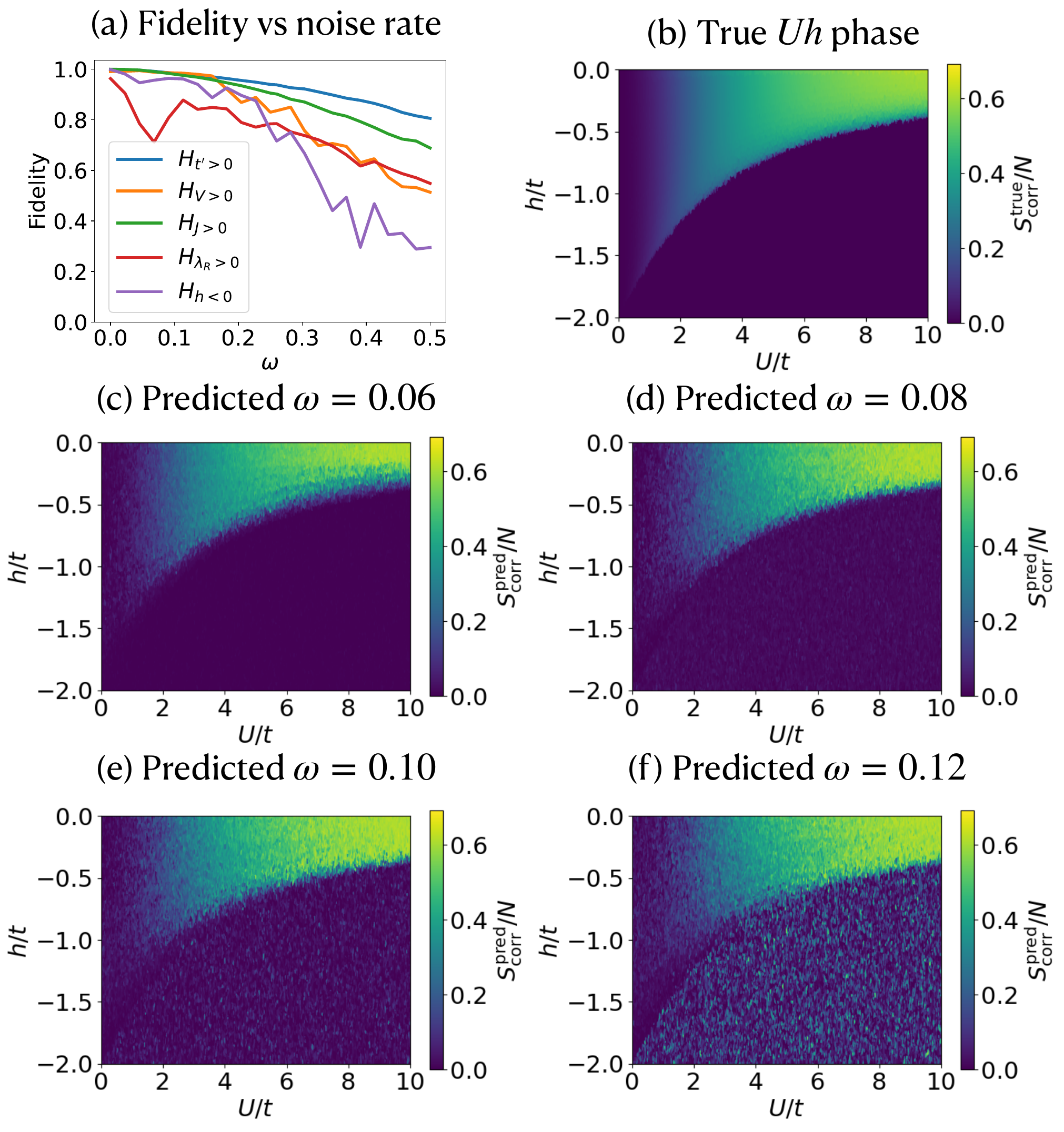}%
\caption{\label{fig:noise} (a) The fidelity of the prediction of each dataset as a function of the noise rate $\omega$. The algorithm is trained on the model $H_{\mu>0}$. (b) True $Uh$ phase diagram. (c-f) Predicted $Uh$ phase diagrams at different noise rate.  }%
\end{figure}

For each value of the noise amplitude, the trained neural network algorithm in a parameter regime is used to predict the correlation entropy of other parameter regimes. The result is shown in Fig.~\ref{fig:noise}. Here, the algorithm is trained on $H_{\mu>0}$, and the fidelity of some predicted data sets is plotted as a function of the noise rate $\omega$. 

We now elaborate on the robustness of our methodology. The transfer learning approach can successfully predict correlation entropy across different quantum phases, not just within the same phase. We note that our algorithm was trained on a single minimal model, and is evaluated in all the other different models. This includes transitions between phases, as long as the underlying quantum orders share certain similarities. For example, we demonstrated that an algorithm trained on one extended Hubbard model can yield reliable predictions in another extended Hubbard model, provided the phases are not drastically different in their quantum order.

However, the method does have limitations: it fails to generalize to phases with fundamentally different quantum orders from the training set (e.g., trying to transfer from antiferromagnetic to ferromagnetic phases, or to those involving Rashba spin-orbit coupling). 
We emphasize that our goal is not merely to identify identical phases across different variants of the extended Hubbard model, but rather to explore how well the model generalizes whether it can predict even previously unseen phases that share similar underlying (antiferromagnetic) quantum orders, such as the bond-order wave phase.

We also emphasize that even when the transfer learning method fails to accurately predict correlation entropy for phases that are not included in the training data, it still provides valuable qualitative insights. Specifically, we observe that the model's prediction uncertainty (standard deviation) is significantly higher for these "unseen" phases. This large variance effectively acts as a signal, highlighting regions where the model encounters unfamiliar quantum orders. Consequently, this feature helps distinguish between phases where the model is reliable and those where it is not, providing an indirect indication of phase boundaries. This characteristic demonstrates that our approach not only predicts known phases accurately but also identifies areas that may host novel or qualitatively different phases.

As expected, the fidelity of all the predictions decreases as the noise rate increases. However, datasets with an antiferromagnetic nature are more robust to noise, especially for smaller noise. The $Uh$ phase at different noise rates is shown as examples in Fig.~\ref{fig:noise}(c-f). The prediction of the correlation entropy of a system with Rashba spin-orbit coupling on the other hand, is largely affected by random noise and fluctuates for different noise rates. This is expected as the trained data set $H_{\mu>0}$ does not include similar quantum orders.  

\section{Conclusion\label{sec:conclusion}}

In this work, we show that machine learning algorithms trained to extract correlation entropy can generalize to different model regimes, provided they exhibit similar quantum orders.
Our methodology relies on 
deep neural network algorithms to predict quantum correlation entropy from local measurements in interacting fermionic systems. By incorporating transfer learning, we demonstrated
the ability to predict correlation entropy across variants
of Hamiltonians not considered in the training set.
Our results demonstrate that neural network algorithms trained on systems with similar quantum order can accurately predict correlation entropy in regimes with different
microscopic parameters, even when parent Hamiltonians are widely different. 
We also showed that in the selected cases where transfer learning fails,
such failure can be automatically detected from the machine learning models alone
by using the prediction fluctuation, 
and does not require knowledge of the real correlation entropy.
Furthermore, the prediction fluctuation allows to directly
signal phase transitions in the quantum many-body phase diagram,
even when those phases were never considered in the training. 
We showed that our algorithm is robust to moderate levels of noise in the input correlators without significant loss in prediction accuracy. {Our approach not only predicts correlation entropy accurately but also identifies unfamiliar phases through large fluctuations among different models, providing a direct signal of phase transitions even without prior knowledge of those phases.} Our transfer learning approach shows that correlation entropy learning
could be performed experimentally using selected local measurements, 
even in regimes where systems feature Hamiltonians
that have not been considered in the training. 

\textbf{Acknowledgements:}
We acknowledge financial support from the Finnish Quantum Flagship.
T.O. acknowledges the Academy of Finland Project No. 331094 for support. 
J.L.L. acknowledges financial support from
the Academy of Finland Projects Nos. 331342, and 358088,
InstituteQ, the Jane and Aatos Erkko Foundation
F.A. and J.L.L. acknowledge the
computational resources provided by the Aalto Science-IT
project.

\appendix

\section{Machine learning methodology and data generation}

\begin{table}[b]
    \centering
    \begin{tabular}{c|c}
         Input Data & \makecell{$\langle c_{is}^{\dagger}c_{js'}\rangle$, $\langle n_{is}n_{js'}\rangle$} \\ [0.3em]
         \hline \\ [-0.9em]
         Optimizer & Adam \\ [0.3em]
         \hline \\ [-0.9em]
         Learning Rate & Decaying with 2.5\% each epoch\\ [0.3em]
         \hline \\ [-0.9em]
         Structure &  Dense$_{(1024)}$/Dense$_{(1024)}$/Dense$_{(1024)}$  \\ [0.3em]
         \hline \\ [-0.9em]
         Target data & $S_{\rm{corr}}/N$ \\ [0.3em]
         \hline \\ [-0.9em]
         Training time & \makecell{2-3 hours, CPU type:\\ Intel Xeon Gold 6148 (Skylake, AVX-512)}
    \end{tabular}
    \caption{Summary of the structure of the neural network algorithms.}
    \label{tab:structure_neural_network}
\end{table}

The correlators used as input for training the correlation entropy for each extended Hubbard model in Eq.~\eqref{eq:Hamiltonians} are obtained by solving the corresponding Hamiltonians separately. The parameter ranges used are: $V \in [0, 8]$ for $H_{V>0}$, $t' \in [0, 1]$ for $H_{t'>0}$, $\lambda_R \in [0, 1.2]$ for $H_{\lambda_R > 0}$, $J \in [0, 1]$ for $H_{J > 0}$, $\mu \in [0, 1]$ for $H_{\mu > 0}$, and $h \in [0, 1]$ for $H_{h > 0}$. For negative-valued parameters, the corresponding maxima are reflected to minima via a minus sign. This is done using the tensor network formalism~\cite{PhysRevLett.69.2863,2020arXiv200714822F,ITensor,DMRGpyLibrary} with open boundary conditions on a one-dimensional fermionic chain with 32 spinful sites. The correlators are extracted from the four central sites in the chain. For particle-particle correlators in Eq.~\eqref{eq:ppcorrelator}, 32 correlators are extracted from the four sites. For density-density correlators, 32 correlators are extracted from $\left\langle n_{is}n_{js'}\right\rangle$ to form the correlators in Eq.~\eqref{eq:nncorrelator2}. A summary of the details of the neural-network algorithm is illustrated in table \ref{tab:structure_neural_network}. For the 32-site chain, the number of correlators needed to determine the correlation entropy is about $1.5\%$ of the entire correlators in the correlation matrix. For larger chains, this percentage decreases because the correlation entropy becomes independent of the system size.~\cite{10.21468/SciPostPhysCore.6.2.030}. 

\begin{figure}[t!]
\includegraphics[width=1.0\linewidth]{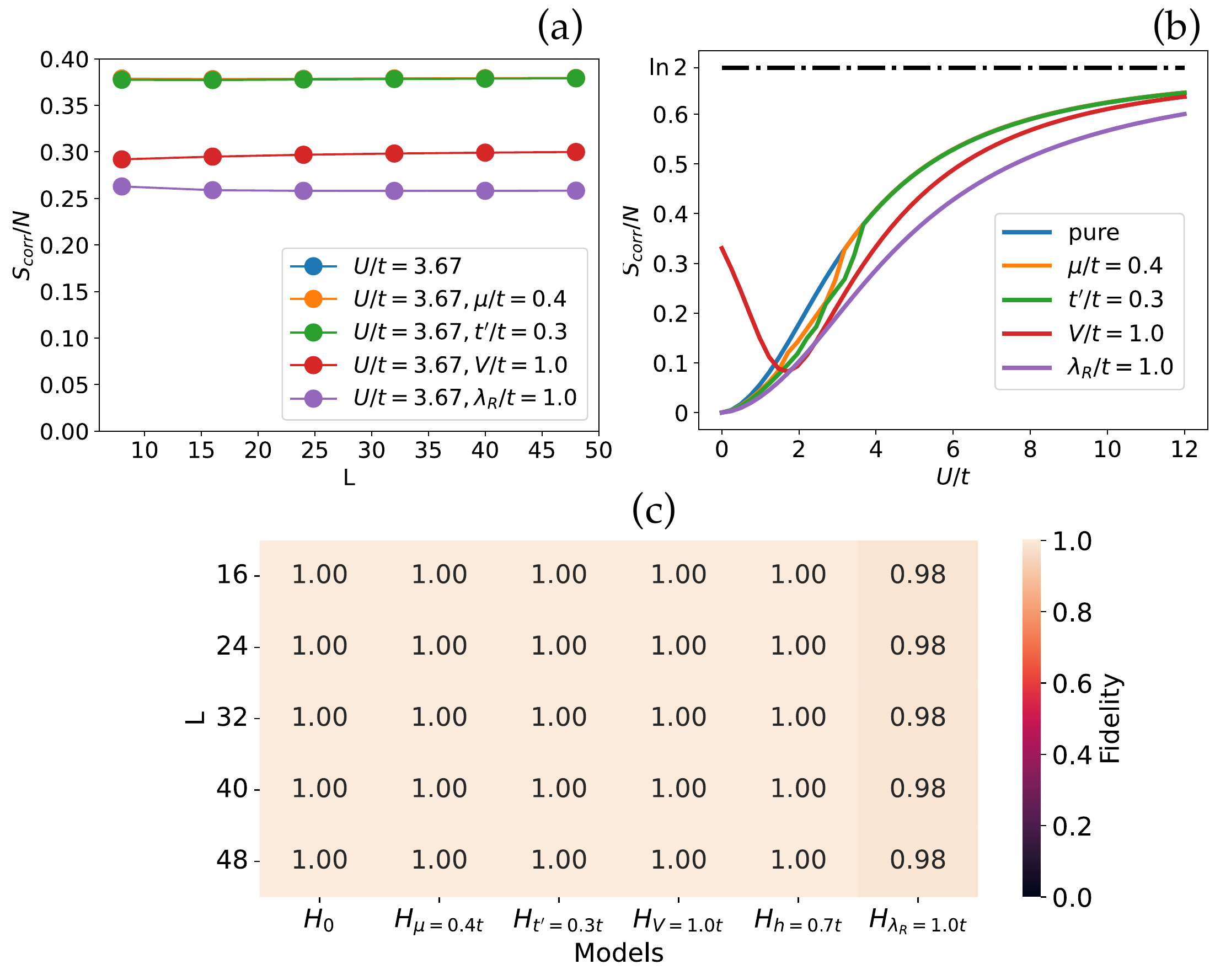}%
\caption{\label{fig:generalizability} (a) Size independence of the correlation entropy at $U=3.67t$. (b) Dependence of the correlation entropy of a 32-site chain on on-site interaction strength $U$. The dashed black curve is where $S_{\rm{corr}}/N=\ln 2$. (c) Fidelity of prediction trained on data obtained from 32-site $H_{h>0}$ to predict on same dependence as (b) for different chain sizes. }%
\end{figure}

We now briefly elaborate the strategies to optimize the performance of our algorithm. First, for data preparation, solely using the particle-particle correlators is not enough to predict quantum phases that are not included in the training process, and hence they must be combined with density-density correlators. Second, during the training process, we employ a decaying learning rate that decreases by 2.5\% with each epoch, enhancing convergence. A decaying learning rate can also avoid overfitting and is beneficial for generalization.
With regards to optimizers, we experimented with different optimizers and finally used Adam for generating the best result. 
The structure of the each neural network algorithm is same as shown in Fig.~\ref{fig:schematics}(a). The input correlators are connected into three hidden layers, each containing 1024 nodes. Each training takes about 2-3 hours to complete. A summary of the structure of the neural network algorithms is shown in Fig.~\ref{tab:structure_neural_network}.
The data and codes implementing the data generation and training are available in Zenodo~\cite{zenodolink}.

\section{Generalizability of the neural-network algorithms}
\label{sec:Generalizability}

\begin{figure}[t]
\includegraphics[width=1.0\linewidth]{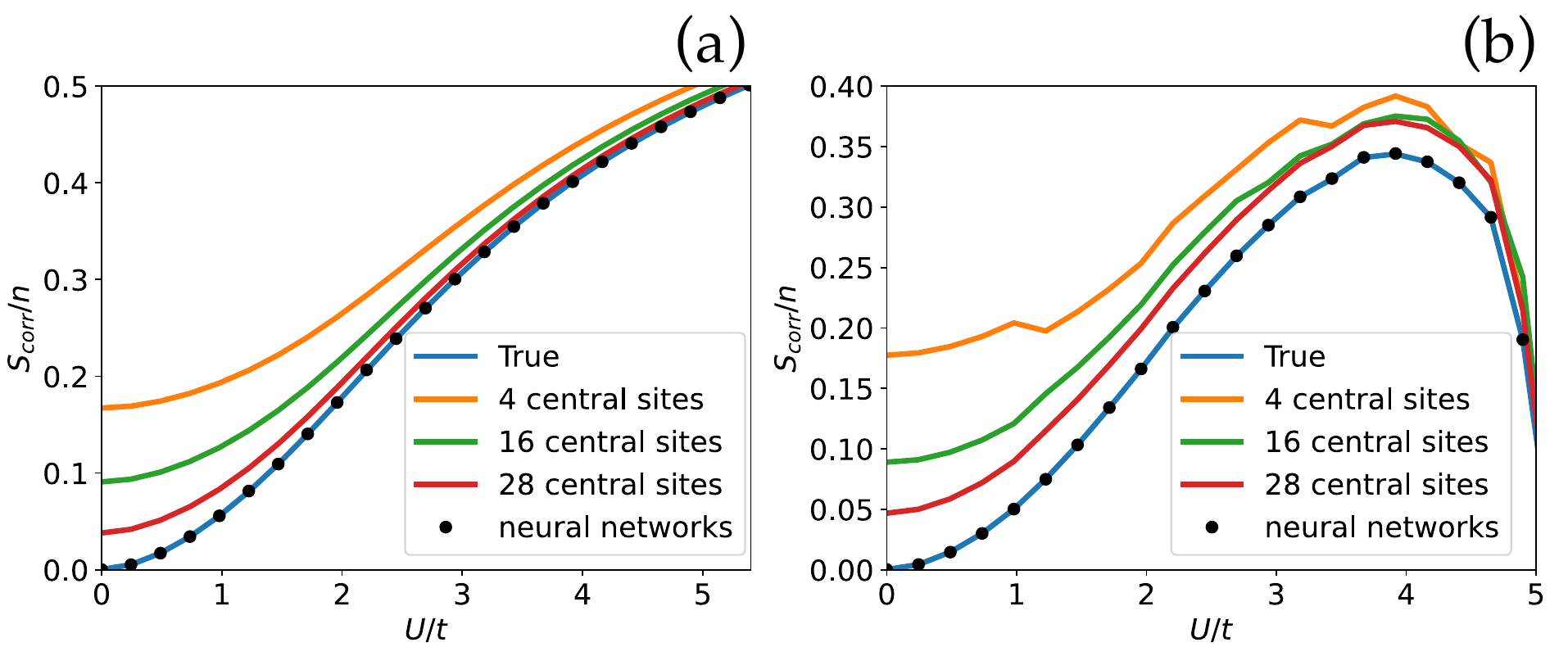}%
\caption{\label{fig:block_method} Estimating the correlatin entropy by the restricted block approach from the reduced correlation matrix for (a) pure Hubbbard model and (b) an extended Hubbard model with a magnetic field in $z$ direction of a 32-site chain. Here the correlation entropy is normalized to the number of sites $n$ related to the correlators in the central blocks. }%
\end{figure}

The Hubbard interaction $U$ as the main drive of the correlation entropy $S_{\rm{corr}}$. Hence it is instructive to examine the dependence of $S_{\rm{corr}}$ for various extension of the spinful interacting fermionic chain, as shown in Eq.~\eqref{eq:Hamiltonians}. The results are shown in Fig.~\ref{fig:generalizability}(b). The correlation entropy per site vanishes for quantum states that do not support sizable entanglement and saturates to the value $\ln(2)$ otherwise. This can also be seen analytically in a pure Hubbard dimer. In this case the ground state solution can be written as
\begin{equation}
    \vert \Psi_0 \rangle = \frac{\left[c_{1\uparrow}^{\dagger}c^{\dagger}_{1\downarrow}+c_{2\downarrow}^{\dagger}c_{2\uparrow}^{\dagger}+\alpha\left( c_{1\uparrow}^{\dagger}c_{2\downarrow}^{\dagger}-c_{1\downarrow}^{\dagger}c_{2\uparrow}^{\dagger}\right) \right] }{2\sqrt{1+\alpha W}}\vert \Omega \rangle,
\end{equation}
where $\vert \Omega \rangle$ is the many-body empty state, $\alpha=W+\sqrt{1+W}$ and $W=U/(4t)$. Then one can construct the correlation entropy analytically and obtain the correlation entropy as
\begin{equation}
    S_{\rm{corr}}=2\ln\left(\frac{2\sqrt{1+W^2}}{W} \right)-\frac{2\ln\left(\frac{1+\sqrt{1+W^2}}{W} \right)}{\sqrt{1+W^2}}.
\end{equation}
Then for $S_{\rm{corr}}/N \xrightarrow{U\rightarrow\infty}\ln2$, where $N=2$. 

As shown in Ref.~[\onlinecite{10.21468/SciPostPhysCore.6.2.030}], the correlation entropy already becomes independent of system size for 32 site model. Figure \ref{fig:generalizability}(a) further demonstrates this size-independent behavior across various extended versions of the Hamiltonian defined in Eq.~\eqref{eq:Hamiltonians}. Thus, since the correlation entropy is independent of system size, the neural network algorithms used in this work are also size-independent. Consequently, the fidelity values presented in Fig.~\ref{fig:fidelity}(a) can be directly extended to systems of different sizes. Figure \ref{fig:generalizability}(c) shows the performance of a neural-network algorithm trained on data from the 32-site $H_{h>0}$, predicted on data exhibiting the same dependence as in Fig.~\ref{fig:generalizability}(b), across different system sizes. The fidelity values closely match those in the fidelity table shown in Fig.~\ref{fig:fidelity}(a), even across different system sizes. 

\begin{figure}[t]
\includegraphics[width=1.0\linewidth]{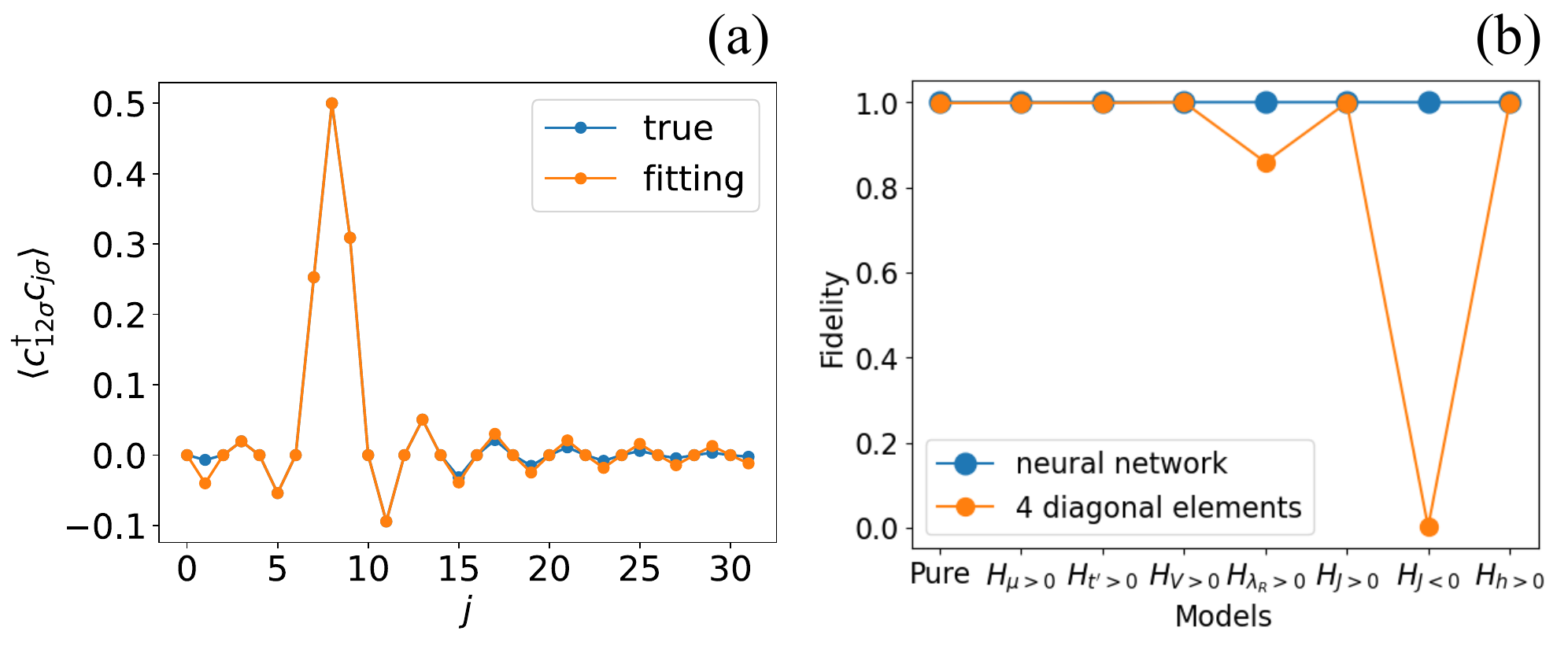}%
\caption{\label{fig:alternative} (a) Fitting the correlation matrix of a 32-site Hubbard model by using Eq.~\eqref{eq:alternative} of a pure Hubbard model, where the fitting is done for 12th row of the correlation matrix at $U=2.7t$. (b) A comparison between the fidelity values of neural-network algorithm and the fitting method of a 32-site Hubbard model.}%
\end{figure}

\section{Limitations of Central Block Approaches to Correlation Entropy}

As a baseline, we attempted to estimate the correlation entropy $S_{\rm corr}$ directly from the central block of the one-particle correlation matrix, by computing $S(C_{\rm block})/n_{\rm block}$ for increasing block sizes $L_{\rm block}$. As shown in Fig.~\ref{fig:block_method}, we found that this naive approximation fails to reproduce the correlation entropy with satisfactory accuracy. While increasing the block size improves the estimate, the number of correlators required already far exceeds those used in the neural network approach. Furthermore, this method is strongly system size dependent. Larger chains demand an increasing number of correlators to achieve comparable accuracy. Such inaccuracies arise from truncating the correlation matrix by discarding long-range correlations, which are crucial for an accurate assessment of many-body entanglement. This failure highlights the necessity of the neural network approach, which can reliably infer the correlation entropy from only a small set of local measurements.

\section{Alternative approach for determining the correlation entropy}

\begin{table}[t]
    \centering
    \begin{tabular}{c|c|c}
         Feature & \makecell{Neural-network} & Fitting method\\ [0.3em] 
         \hline \\ [-0.9em]
         \makecell{System size\\ independence} & Yes & No \\ [0.3em]
         \hline \\ [-0.9em]
         \makecell{Required\\ number of\\ correlators} & 64 & \makecell{$\sum_{i=0}^{d-1}(2N-i)$\\$=d(4n-d+1)/2$,\\ where $d$ is the \\number of fitted\\ diagonals of the \\correlation matrix}\\ [0.3em]
         \hline \\ [-0.9em]
         Applicability & All & \makecell{Antiferromagnetic\\ systems}\\
         \hline \\ [-0.9em]
         Generalizability & Yes & No\\ [0.3em]
         \hline \\ [-0.9em]
         Transferability & \makecell{Antiferromagnetic\\ systems} & No \\[0.3em]
         \hline \\ [-0.9em]
         \makecell{Number of \\parameters} & \makecell{$\sum_{i=0}^{L} \big( n_i \cdot n_{i+1}$ \\ $+ n_{i+1} \big)$, where $n_i$ \\
         is the number nodes\\
         in each layer, and $i$ \\
         runs over the \\
         number of layers} & 3
    \end{tabular}
    \caption{Summary of the comparison between fitting method and the neural network method.}
    \label{tab:difference_fit_nn}
\end{table}

In certain cases, the correlation entropy can also be obtained through fitting techniques. In this section, we introduce a simple fitting approach that estimates long-range correlators from short-range correlators using a decaying cosine function,
\begin{equation}\label{eq:alternative}
    C_{i,j}=c_0\cos(c_1|i-j|+c_2)/|i-j|,
\end{equation}
where $c_0$, $c_1$, and $c_2$ are determined numerically for each row of the correlation matrix. An example is shown in Fig.~\ref{fig:alternative}(a). Note that this fitting method requires prior knowledge of the $n$ diagonal elements of the correlation matrix, and hence it is also system size dependent. Furthermore, this fitting approach can only be applied when the fitted and predicted correlators share the same site index $i$ in $C_{ij}$. It cannot be used to infer other short-range or long-range correlators that are not directly related to the fitted ones (i.e., when $i$ differs between the fitted and predicted correlators). Consequently, Eq.~\eqref{eq:alternative} is insufficient to reconstruct the entire correlation matrix solely from the central four sites. In addition, edge correlators, which are strongly dependent on system size, cannot be reliably captured by this fitting method. In Fig.~\ref{fig:alternative}(b), we present the fidelity values of this fitting method fitted with four diagonal elements applied to various extended versions of the 32-site Hubbard model. For comparison, the figure also includes the corresponding predictions obtained from the neural network method. The results show that while the fitting method performs well in certain cases, it fails to accurately reproduce the correlation entropy for the extended Hubbard model with Rashba spin–orbit interaction and ferromagnetic spin–spin interaction. The comparison between two methods are summarized in Table \ref{tab:difference_fit_nn}. The comparison highlights that the neural-network method is system-size independent, require fewer correlators, and offer broader applicability, generalizability, and transferability than the fitting method, which is limited to antiferromagnetic systems and scales with system size.

\bibliography{apssamp}{}

\end{document}